# PieceTimer: A Holistic Timing Analysis Framework Considering Setup/Hold Time Interdependency Using A Piecewise Model

Grace Li Zhang, Bing Li, Ulf Schlichtmann
Institute for Electronic Design Automation, Technical University of Munich (TUM), Germany
{grace-li.zhang, b.li, ulf.schlichtmann@tum.de}

## ABSTRACT

In static timing analysis, clock-to-q delays of flip-flops are considered as constants. Setup times and hold times are characterized separately and also used as constants. The characterized delays, setup times and hold times, are applied in timing analysis independently to verify the performance of circuits. In reality, however, clock-to-q delays of flip-flops depend on both setup and hold times. Instead of being constants, these delays change with respect to different setup/hold time combinations. Consequently, the simple abstraction of setup/hold times and constant clock-to-q delays introduces inaccuracy in timing analysis. In this paper, we propose a holistic method to consider the relation between clock-to-q delays and setup/hold time combinations with a piecewise linear model. The result is more accurate than that of traditional timing analysis, and the incorporation of the interdependency between clock-to-q delays, setup times and hold times may also improve circuit performance.

## 1 Introduction

Modern IC design faces tremendous challenges in maintaining a continuously improving performance. To achieve this goal, new manufacturing technologies have been introduced every several years to improve the performance of circuit components, and the circuit design flow has been refined relentlessly to meet the requirements of new technologies and design methodologies.

In the advance of the semiconductor industry, the design flow follows the manufacturing technology closely to bring the new progress into the design domain. In adopting a new manufacturing technology, the design flow abstracts lower level details and exposes only the extracted device and circuit models of different levels to designers, so that the huge logic resource become manageable in the design domain.

As a representative of the abstraction at circuit level, static timing analysis (STA) assumes that combinational circuit stages between flip-flops are independent, and a flip-flop can work reliably if its setup time and hold time are met [1, 2]. With this assumption, only the worst/best corners (late and early modes) need to be verified with the largest/smallest combinational delays. Consequently, manufacturing details are hidden behind the abstracted gate delay models, and the performance of the circuit is evaluated as independent of applications.

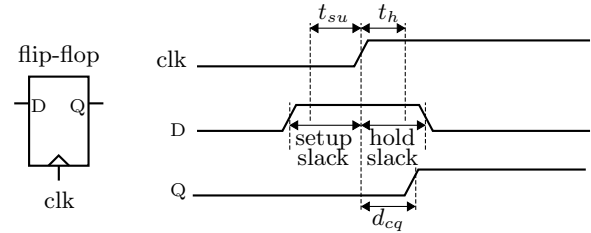

**Figure 1: Setup time ($t_{su}$), hold time ($t_h$) and clock-to-q delay($d_{cq}$) of a flip-flop.** *Setup slack and hold slack* **are defined as the distance from the latest and earliest signal switching to the active clock edge, respectively.**

In STA, a flip-flop is characterized by its setup time, hold time, and clock-to-q delay. These characteristics can be explained using the example in Figure 1. If the latest signal switching at the input of the flip-flop is $t_{su}$ time before it is sampled by the active clock edge, and if the earliest signal switching happens $t_h$ time later than the active clock edge, the data at the input of the flip-flop can be latched correctly. In this case, the delay from the active clock edge to the output of the flip-flop (Q), clock-to-q delay $d_{cq}$, is characterized as a constant. If either of the two constraints is violated, the flip-flop is considered to not work properly, and a timing error is reported.

The timing model of a flip-flop with setup time, hold time and constant clock-to-q delay described above is a significant simplification for timing analysis and optimization. As a result, the combinational circuits between flip-flops can be analyzed and optimized independently without considering time borrowing as in designs using level-sensitive latches. This efficiency in timing analysis is very important because front-end circuit design usually undergoes many analysis-optimization iterations.

The simplified flip-flop model above, however, sacrifices circuit performance for the sake of execution efficiency. In order to simplify the clock-to-q delay to be a constant, the latest signal switching must be $t_{su}$ earlier than the clock edge. Otherwise, the circuit is not considered to work properly. In reality, however, the flip-flop may still latch the input data correctly if the arrival time of a signal is late, although the clock-to-q delay may become larger than the constant delay. If the arrival time of the input signal is too close the clock edge, the flip-flop may finally enter metastability and thus fail to work properly. Similarly, if the hold time constraint is violated, the flip-flop also works with a larger delay in a feasible region, before the signal change is too close to the clock edge. For a given signal, we refer to the distance from the signal switching at the input of a flip-flop to the clock edge as *setup slack*, and the distance from the clock edge to the signal switching as *hold slack*, as illustrated in Figure 1. Consequently, setup time and hold time in the traditional definition are actually also slacks at which the clock-to-q delay of the flip-flop is characterized as a constant.

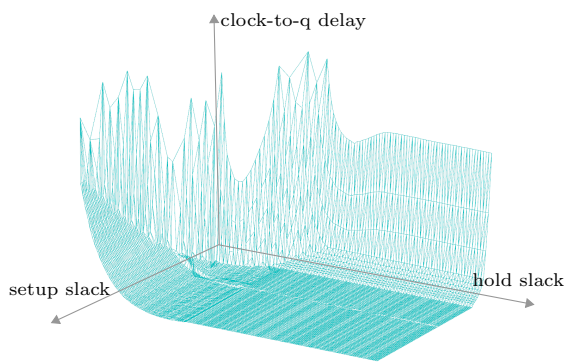

**Figure 2: The three dimensional clock-to-q delay surface of a 45nm flip-flop with respect to setup slack and hold slack.**

The relation between clock-to-q delay, setup slack and hold slack can be demonstrated using the simulated delay surface of a 45nm flip-flop in Figure 2. In this example, we can see that when the setup and hold slacks are large, the delay of the flip-flop is a constant. If these slacks are reduced enough, the clock-to-q delay becomes larger, until the flip-flop finally enters metastability. In traditional STA, the flip-flop is assumed to work in the area with a constant clock-to-q delay. This simplification does not take advantage of the feasible region beyond the setup time and hold time, so that the circuit performance may be underestimated.

If the flip-flop is allowed to work in the region with the setup slack smaller than setup time, the clock period for the critical path of the circuit can be smaller, because the latest signal switching can arrive at the ending flop-flip later. Accordingly, the delay of this flip-flop becomes larger, but the increased delay only affects the combinational path in the next stage. In case the delay of the combinational path of the next stage is not large, no timing violation appears. Consequently, the critical path receives more timing budget for signal propagation, leading to an improved circuit performance. This delay compensation is very similar to designs with level-sensitive latches, or designs using intentional clock skew scheduling. The advantage of this phenomenon is that the performance increase comes from a more accurate timing analysis, without invoking the complete design-optimization flow or timing ECO. Therefore, timing analysis considering this delay compensation is specially useful in late stages of design flow such as timing signoff.

To take advantage of the relation between clock-to-q and setup/hold slacks, several methods have been proposed. The method in [3, 4] exploits the compensation between setup slack and hold slack with respect to a given clock-to-q delay and produces results with more relaxed slacks. To reduce simulation time, a method based on Euler-Newton tracing to characterize this curve with respect to setup/hold slacks efficiently is proposed in [5, 6]. These methods only consider the relation between setup slack and hold slack, so that the three dimensional interdependency problem is simplified into a two-dimensional problem.

To consider the three dimensional interdependency between clock-to-q delay, setup slack and hold slack, the method in [7] uses a quadratic programming model to calculate the optimal clock period directly, but it is limited due to the scalability of this high-order programming method. To simplify the three dimensional model, the method in [8] approximates the relation between clock-to-q delay and setup/hold slacks using an analytic function and calculates the minimum clock period of a circuit by iterations. This method, however, cannot guarantee to converge in the given number of iterations. In addition, the method in [9] approximates the three dimensional delay surface using linear planes, but in calculating the minimum clock period this method splits the problem into two dimensional problems, so that it cannot guarantee an optimal solution. Furthermore, the method in [10] proposes a very efficient algorithm to capture timing violations in a circuit, but it only considers the relation between clock-to-q delay and setup slack.

In this paper, we propose a holistic method to characterize the three dimensional delay surface and calculate the minimum clock period using a piecewise ILP model. Our contributions are as follows:

- We characterize the three dimensional delay surface by approximating it with small polygons, either triangular or rectangular. For each polygon, we only simulate the real flip-flop delays using SPICE at its corners, and verify whether the linear plane defined by these corners can approximate the real clock-to-q delay with a given accuracy. Since not all the delay values inside a polygon are simulated, this method can reduce the characterization time significantly. In addition, this piecewise model has the advantage to approximate nearly any surface with a sufficient number of small polygons.

- The piecewise delay surface is used to calculate the minimum clock period considering the interdependency between clock-to-q delay and setup/hold slacks. Since this is the first time that the interdependency is considered in the three dimensional space directly with enough modeling accuracy and scalability, the proposed method evaluates the clock period more accurately than previous methods and timing violations can be removed effectively.

The rest of this paper is organized as follows. In Section 2 we review the background and the state-of-the-art research on the interdependency between clock-to-q delay and setup/hold slacks. In Section 3 we explain our method to characterize the three dimensional delay surface of a flip-flop using piecewise polygons. This model is used to calculate the minimum clock period with an ILP formulation. Experimental results are shown in Section 4. The conclusion is given in Section 5.

## 2 Background and state of the art

Timing constraints of digital circuits can be explained using Figure 3, where three flip-flops are connected by combinational circuit blocks. When an active clock edge is generated, all flip-flops are triggered at the same time. In traditional

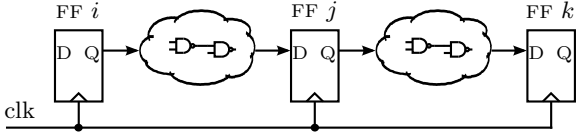

**Figure 3: Circuit example with three flip-flops and two combinational circuits.**

STA, timing constraints are defined with setup time, hold time and constant clock-to-q delays. For flip-flops $i$ and $j$, the timing constraints can be written as

$$d_{cq,i} + \overline{d}_{ij} + t_{su,j} \leq T \quad (1)$$
$$d_{cq,i} + \underline{d}_{ij} \geq t_{h,j} \quad (2)$$

where $d_{cq,i}$ is the delay of flip-flop $i$, $\overline{d}_{ij}$ ($\underline{d}_{ij}$) is the maximum (minimum) delay of the combinational circuit between flip-flops $i$ and $j$, $t_{su,j}$ ($t_{h,j}$) is the setup (hold) time of flip-flop $j$, and $T$ is the clock period.

The traditional setup time of a flip-flop is characterized by moving the signal switching at the input of the flip-flop to the active clock edge gradually. As the signal switching approaches the active clock edge, the delay of the flip-flop starts to increase. When this delay reaches a given metric, e.g., 110% of the minimum delay of the flip-flop characterized with very large setup and hold slacks, the time difference between the signal switching and the active clock edge is defined as the setup time. Hold time is characterized similarly by moving the signal switching after the active clock edge to the clock edge gradually. Accordingly, this increased delay of the flip-flop, 110% delay in this case, is used as the clock-to-q delay in timing analysis.

The characterization process above guarantees that the circuit works properly if the constraints (1) and (2) are met for each pair of flip-flops, because the reserved setup time from latest signal switching to the active clock edge and the reserved hold time from the clock edge to the earliest signal switching together guarantee that the clock-to-q delay of the flip-flop does not exceed the characterized delay. This guarantees further that the timing constraints of the next flip-flop stage, e.g., between flip-flop $j$ and $k$ in Figure 3 can be verified by assigning the clock-to-q delay of the flip-flop $j$ to the characterized constant delay. In this way, the timing dependency between consecutive flip-flop stages is hidden from designers.

The timing characterization above simplifies the delay model of a flip-flop. As explained earlier, when setup slack or hold slack becomes smaller, the delay of the flip-flop increases. Figure 4 illustrates several curves with respect to the setup slack and hold slack as in [9]. On each curve, the clock-to-q delay is a constant, and the curve closer to the axes has a larger delay. Before the delay becomes very large and soon the flip-flop enters the metastable region, all these curves are valid and the flip-flop can work with each setup/hold slack combination on them. The traditional setup/hold time delay model, however, only assumes that the flip-flop works with curves with a delay no larger than the characterized delay, e.g., 110% of the stable delay. All the other curves on the left of the 110% curve are simply ignored.

When characterizing the traditional setup time, the hold slack is set to a very large value to exclude its influence. The setup slack is then decreased gradually, until the delay of the flip-flop increases to 110%. The hold time is characterized similarly. Therefore, the setup and hold time point used for static timing analysis is shown as point A in Figure 4b. When point A is used in STA, the shaded area under the curve is also included. This area is located below the curve, so that the delay in the area is larger than 110% delay. However, the flip-flop delay used in STA is still assumed as the characterized 110% delay, posing a risk that the circuit still does not work even all timing constraints are met. To solve this problem, the method in [3, 4] verifies whether all the setup and hold slack pairs in the circuit are above the 110% curve, meaning the delay of all the flip-flops are smaller than the 110% delay.

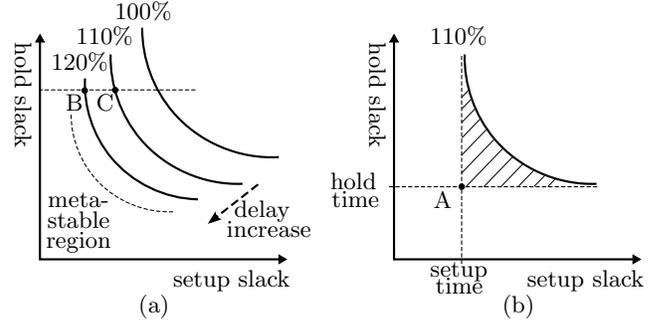

**Figure 4: Delay curves of a flip-flop. (a) Curves of setup/hold slack combinations with respect to different constant clock-to-q delays. (b) Characterization point of setup time and hold time in traditional STA.**

The method in [3, 4] excludes the delay curves lower than the 110% delay curve in timing analysis, so that the clock period cannot be improved. For example, assume flip-flop $j$ in Figure 3 works with a small setup slack, so that the clock period for this flip-flop stage can be lowered. This small setup slack may incur an increase of the clock-to-q delay of flip-flop $j$. This delay increase, however, may be absorbed by the stage between flip-flops $j$ and $k$, if this stage has a small combinational delay, so that the lowered clock period works with the whole circuit. This scenario is similar to the case that flip-flop $j$ works at the point B in Figure 4a and flip-flop $k$ still works at point C.

The delay compensation case has been investigated by the methods in [8, 9, 10]. The method in [10] considers the flexible clock-to-q delay in timing analysis, but without modeling the joint effect of setup and hold slacks. In this method, a flip-flop is allowed to work with different delays due to setup slacks, similar to the points on different curves at a given hold slack. Consequently, the interdependency between the clock-to-q delay and the setup time slack becomes a two dimensional problem, resulting in a highly efficient timing analysis algorithm. The method in [9] models the three dimensional delay surface directly, and solves the delay compensation problem with respect to setup time slack and hold slack separately, actually also transforming the three dimensional problem into two-dimensional. The method in [8] approximates the delay surface using an analytic model, but the proposed iteration-based method cannot guarantee to converge in a given number of iterations.

As described above, the previous methods either cannot solve the STA problem considering the interdependency between clock-to-q delay, setup slack and hold slack, or cannot guarantee the quality of the solution. In this paper, we propose a holistic method to calculate the minimum clock period of a circuit by modeling the three dimensional delay surface using piecewise polygons. Thereafter, we calculate the minimum clock period using a piecewise ILP model combined with trimming techniques. The proposed method is very efficient in both delay modeling and timing analysis.

## 3 Piecewise linear delay model and timing analysis

Static timing analysis considering the flexibility of flip-flop delays needs to solve two problems: 1) The three dimensional surface of the flip-flop should be modeled. In this step, only the necessary delay information that can be used to calculate the minimum clock period should be retained. Other delay information should be omitted to reduce simulation time; 2) A timing analysis algorithm using the piecewise delay model to calculate an accurate minimum clock period for a circuit. This algorithm should take the delay compensation across flip-flop stages into account and calculate the minimum clock period in a reasonable time.

### 3.1 Adaptive piecewise polygonization of a three dimensional delay surface

To transform the delay surface in Figure 2 into a form that can be used by static timing analysis, we partition it into small regions and approximate each one using a linear plane in the region, or a polygon. This piecewise approximation has the flexibility enabling the tradeoff between runtime and accuracy. The more polygons into which the surface is partitioned, the more accurate this approximation is, but the more time is required to generate these polygons and the slower the timing analysis algorithm becomes.

The linearization of a three dimensional surface is a problem studied widely and many methods have been proposed [11, 12], using techniques such as adaptive sampling. But these methods are very general and do not take advantage of the special shape of the three dimensional delay surface as shown in Figure 2. In this section, we propose an adaptive method to approximate the delay surface using a set of polygons. This piecewise delay model can be used by the timing analysis algorithm in Section 3.2 directly.

The linearization of the delay surface includes three steps. First, we identify the boundary of the surface projected to the setup/hold slack dimensions using triangles. When the setup slack or the hold slack is very small, the flip-flop enters the metastable region, which should be excluded from the valid working space of the flip-flop. Second, we partition the delay surface inside the boundary with rectangles. We then check the approximation accuracy of each rectangle and split it further if the approximated delay is too far away from the real delay on the surface. Third, we merge the rectangles produced in the second step so that the number of polygons can be reduced, which improves the efficiency of timing analysis using this piecewise model.

#### 3.1.1 Approximating the surface boundary using triangles

When the setup slack or the hold slack becomes too small, the flip-flop may enter the metastable region. The boundary between the working region of a flip-flop and its metastable region has a shape similar to the curve in Figure 5a when projected into the setup and hold dimensions.

In the proposed method, the boundary of the delay surface is approximated by a chain of linear segments whose ending points are on the boundary curve, as shown in Figure 5a. As we do not know the shape of the curve, we generate these linear segments to approximate it. We start from a single linear segment shown as the right-most segment connecting points A and B where the setup slack and hold slack are set to large values, respectively.

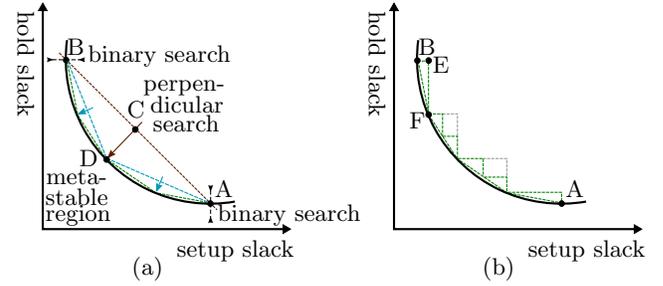

Figure 5: Approximation of the boundary of the delay surface. (a) Approximation with linear segments. (b) Delay surface approximation with triangles.

In this setting, the setup slack at point A does not affect the delay of the flip-flop, so that a binary search of different hold time slacks can capture the delay value at A quickly. In each search, we run SPICE simulation at the target point. If the clock-to-q delay is larger than a given threshold or the SPICE simulator does not return a valid delay, this point is considered outside the feasible region of the clock-to-q delay. Otherwise, we find a valid point in the feasible region. After the binary search is finished, the valid point with the smallest hold slack is identified as point A in Figure 5a. Similar to this process, we assign the flip-flop a very large hold time slack and execute a binary search along the setup slack to identify point B.

The linear segment connecting A and B (A↔B) still cannot be used as an approximation of the surface boundary, because the point at the middle of this segment may be far away from the real boundary. To check the distance between this approximation point (C) and the real boundary point (D), we apply a binary search in the direction perpendicular to the segment A↔B toward D using SPICE simulation. Afterwards, we compare the distance between C and D. If it is larger than a given threshold ($k_{th}$), we split the segment A↔B and create two linear segments A↔D and B↔D. In this way, the distance between the linear segments and the real boundary is reduced. This refining process is repeated further until the entire boundary curve is approximated within the accuracy requirement.

With the linear segments following the boundary closely, we then create triangles to approximate the three dimensional delay surface in the area close to this boundary. First we create triangles using these linear segments as their hypotenuse, as shown in Figure 5b. Each of these triangles defines a valid region in which the delay surface is approximated with a linear plane defined by the corner points of the triangle. For example, in the triangle formed by the points B, E, and F, the clock-to-q delays at B and F are already known from SPICE simulation during constructing the linear segments. We then run SPICE simulation at point E. With the coordinates of the three corner points B, E, and F as well as the corresponding clock-to-q delays, the linear plane in the three dimensional space can be constructed easily.

To verify the accuracy of the triangular approximation in the three dimensional space, we select a point inside the triangle and run SPICE simulation again. We then compare the approximated delay by the linear plane and the simulated delay. Since the real delay surface is convex, the point that has an equal distance to the three ending points is chosen for the verification. In a right triangle, this point is at the middle of its hypotenuse.

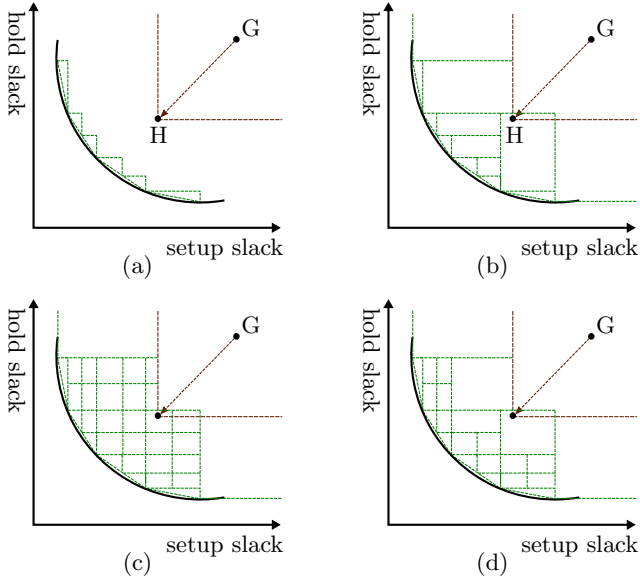

**Figure 6: Rectangle construction, split and merge. (a) Identify the stable plane by search from G to H. (b) Construct rectangle between the triangles and the boundary of the stable region. (c) Split the rectangles to increase modeling accuracy. (d) Merge rectangles to reduce modeling complexity.**

In the comparison above, if the difference between the approximated value and the real delay value is larger than a threshold $d_{th}$, we split the triangle into two as illustrated in Figure 5b. Consequently, we create more linear segments along the delay boundary to increase the approximation accuracy. Note in generating the linear segments along the boundary above, we check the distance of the approximate point to the boundary in the setup/hold slack dimensions using the threshold $k_{th}$. In verifying the approximation accuracy of the delay, we compare the delays directly so that the linear segments approximating the boundary might be split further as shown in Figure 5b.

### 3.1.2 Approximating the delay surface using rectangular polygons

The triangles found above only cover the areas of the delay surface close to the boundary. In order to approximate the whole surface, more polygons are required. Instead of using a lot of polygons directly, we take advantage of the fact that the delay surface is a constant surface when both the setup slack and the hold slack are large. In other words, if the signal switching at the input of a flip-flop is always far way from the active clock edge, the flip-flop always works in a stable region with a constant delay. To approximate this area, we need only one linear plane.

To identify the constant linear plane, we start a search from a stable point (G) at which the setup slack and hold slack are very large, in the direction in which setup and hold slacks decrease at the same time, as illustrated in Figure 6a. At each step, we run SPICE simulation to find the real delay on the surface. Once we reach the first point at which the simulated delay increases, we use the last stable point (H) as the the corner of the stable linear plane.

The ending point H in Figure 6a is hardly the optimal point from which the largest stable region is covered, because there are many such points at which the clock-to-q delay deviates from the stable value on the delay surface. However this non-optimal point does not affect the approximation accuracy or efficiency, because the uncovered region will be split and merged as follows.

To cover the areas between the triangles close to the boundary of the delay surface and the newly identified stable region, we create rectangles and split them with respect to the approximation accuracy. As illustrated in Figure 6b, we start from the edges of the triangles and expand to the right and to the top of the area. Consequently, relatively large rectangles are created. Note there is some overlapping between these rectangles, e.g., the ones covering the corner point H. In the proposed ILP formulation, this overlapping is allowed and the solver chooses one of the points on the overlapping polygons as the working point of the flip-flop. In fact, a working point from any polygon works because they are all delay approximations meeting the specified accuracy.

The corners of a rectangle in Figure 6b are located on the real delay surface directly. Therefore, the largest approximation error likely happens at the center of the rectangle. We run SPICE simulation at each center of the rectangle and compare the real delay with the approximated delay on the plane. If the difference is larger than the threshold $d_{th}$, the rectangle is split further, as shown in Figure 6c.

In forming the rectangles, we simply expand from the triangles. This simplification and the following split may produce more rectangles than necessary. Therefore, we try to merge neighboring rectangles in the last step, because a smaller number of rectangles means fewer constraints in the following ILP formulation. To merge rectangles, we search from each rectangle to the upper and right directions, because the delay in these directions changes relatively slowly. We combine each pair of neighboring rectangles into one rectangle, if the approximation value at the center of the new rectangle is in the range $d_{th}$ from the real delay on the delay surface. The result of the rectangle merging is illustrated in Figure 6d.

### 3.2 Piecewise ILP model for calculating the minimum clock period

After the delay surface is approximated using polygons, we need to calculate the minimum clock period of a circuit. Compared with traditional STA, the challenge of using this piecewise model is to determine on which polygon a flip-flop works. In this section, we describe a method based on ILP formulation to calculate the minimum clock period.

Assume in total there are $n_p$ polygons approximating the delay surface. Since in timing analysis a flip-flop can only work with one setup/hold slack combination, only one of these polygons should be selected for the flip-flop. Therefore, we define a 0-1 variable $z_i^k$ for the $k$th polygon in the piecewise delay model for flip-flop $i$. If the working point of the flip-flop falls into the $k$th polygon, $z_i^k=1$; otherwise, $z_i^k=0$. To allow the solver to choose one and only one polygon, we

specify the following constraint

$$\sum_{k=1,\ldots,n_p} z_i^k = 1. \qquad (3)$$

If the projection of the $k$th polygon to the setup/hold slack dimensions is a rectangle, as illustrated in Figure 7a, the clock-to-q delay of flip-flop $i$ in this region can be expressed as

$$d_{cq,i}^k = f(s_i^k, h_i^k) = c^k \cdot z_i^k + c_s^k \cdot s_i^k + c_h^k \cdot h_i^k \qquad (4)$$

$$s_l^k \cdot z_i^k \leq s_i^k \leq s_u^k \cdot z_i^k, \quad h_l^k \cdot z_i^k \leq h_i^k \leq h_u^k \cdot z_i^k \qquad (5)$$

where $s^k$ and $h^k$ are the setup slack and the hold slack, respectively, and (4) defines the linear plane in the three dimensional delay space.

As described in Section 3.1, the real delays of the flip-flop at the corner points of the rectangle are known from SPICE simulation. Therefore, we can deduce a linear plane which passes the delay points corresponding to the corners of the rectangle. Such a plane can be characterized with only three points, and to be conservative we select three out of the four corner points with the largest delays to create the plane. Consequently, the constant coefficients $c^k$, $c_s^k$, $c_h^k$ in (4) can be determined. Since this polygon is valid only in the rectangular region as illustrated in Figure 7a, the lower and upper bounds of the setup slack are known from the characterization process as $s_l^k$ and $s_u^k$. Similarly, the lower and upper bounds of the hold slack are known as $h_l^k$ and $h_u^k$. In (5) the lower and upper bounds are all multiplied by $z_i^k$ to enable or disable this polygon. If this polygon is selected so that $z_i^k=1$, the constraints (4) and (5) describe a set of linear constraints. If this polygon is not selected with $z_i^k=0$, the ranges of the setup slack and the hold slack are all forced to 0, so that both $s_i^k$ and $h_i^k$ are forced to 0. In this case, the delay $f(s_i^k, h_i^k)$ is also equal to 0, because the constant coefficient $c^k$ is also multiplied by $z_i^k$.

For a triangular region, the corresponding polygon can be defined similar to (4) and (5). To prevent a setup/hold slack combination from falling into the area lower than the hypotenuse of the triangle, we add another constraint as

$$h_i^k \geq c_t^k + c_{t,s}^k \cdot s_i^k \qquad (6)$$

where $c_t^k$ and $c_{t,s}^k$ are characterized constants for the triangle. The concept of this additional constraint can be explained using the example in Figure 7a. The newly added constraint (6) only allows the slack combination to fall into the region above the hypotenuse. Together with (5), this new constraint defines exactly the triangular region.

With the constraints (4)–(6) defined, the setup slack $s_i$, hold slack $h_i$ and the clock-to-q delay $d_{cq}^i$ at flip-flop $i$ can be written as

$$s_i = \sum_{k=1,\ldots,n_p} s_i^k, \quad h_i = \sum_{k=1,\ldots,n_p} h_i^k \qquad (7)$$

$$d_{dq}^i = f(s_i, h_i) = \sum_{k=1,\ldots,n_p} f(s_i^k, h_i^k) \qquad (8)$$

The constraints (7) and (8) are valid because the solver can only select one polygon constrained by (3). Consequently, the slacks and clock-to-q delays $s_i^k$, $h_i^k$ and $d_{cq,i}^k$ can take nonzero values only in one region. Therefore, the sums in (7) and (8) are equal to the slacks and clock-to-q delay at the flip-flop.

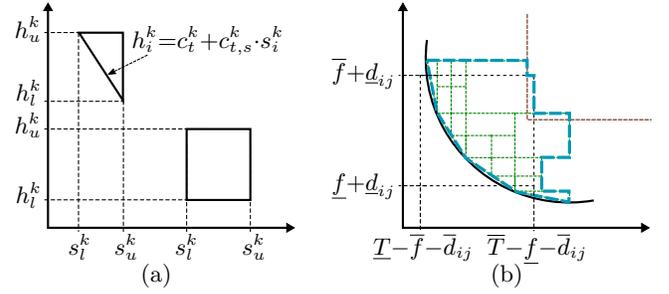

**Figure 7: Slack ranges and polygon trimming. (a) Triangular and rectangular ranges of setup and hold slacks. (b) Trimming polygons using absolute upper and lower bounds of slacks.**

In (1), setup time should be added to the path delay to verify the clock period. This setup time, however, is a virtual metric to guarantee that the clock-to-q delay is no larger than a given value, e.g., 110% of the stable delay. Since we incorporate the increase of clock-to-q delay in the delay model, this setup time is not needed in the constraint anymore. For the same reason, hold time is also removed from the constraint (2). Consequently, the timing constraints between flip-flops $i$ and $j$ can be written as

$$s_j \leq T - f(s_i, h_i) - \overline{d}_{ij} \qquad (9)$$

$$h_j \leq f(s_i, h_i) + \underline{d}_{ij}. \qquad (10)$$

In the constraints (9) and (10), the feasible working region of the flip-flop is already included in the regions of the polygons approximating the delay surface.

With the constraints above, the ILP problem to calculate the minimum clock period can be expressed as

$$\text{minimize} \quad T \qquad (11)$$
$$\text{subject to} \quad (3)\text{–}(10), \forall \text{ flip-flop pair} \qquad (12)$$

Note that we relax the constraints for the variables $s_j$ and $h_j$ in (9) and (10) from equation to inequation to simplify the formulation. Since the values of $s_j$ and $h_j$ returned by the solver are always no larger than the real slacks defined by the right side of (9) and (10), the calculated minimum value of $T$ is always a feasible solution.

Since the interdependency between clock-to-q, setup slack and hold slack allows timing compensation across flip-flop stages, the clock period can be lowered compared with the results of the traditional STA. However, this improvement incurs a large runtime in solving the ILP problem (11)–(12).

To reduce the computational complexity, we trim the polygons in the delay model of flip-flops. The basic idea is that we find absolute lower and upper bounds of the setup slack $s_i$ and the hold slack $h_i$. The polygons completely falling outside the bounding box can be removed from the delay model.

In characterizing the flip-flop delay surface, we know that the delay $f(s_i, h_i)$ are bounded in a range $[\underline{f}, \overline{f}]$, where $\underline{f}$ is the stable delay of the flip-flop when the setup slack and the hold slack are very large, and $\overline{f}$ is the delay beyond which we consider that the flip-flop enters metastability. For the clock period $T$, we also specify its range as $[\underline{T}, \overline{T}]$, where $\underline{T}$ ($\overline{T}$) is the lower (upper) bound of the clock period calculated by setting all setup slacks to the smallest (largest) values from delay characterization, and all clock-to-q delays to the lower (upper) bound $\underline{f}$ ($\overline{f}$). Consequently, we can specify the range of the setup slack of flip-flop $j$ with a combinational path from flip-flop $i$ as $[\underline{T}-\overline{f}-\overline{d}_{ij}, \overline{T}-\underline{f}-\overline{d}_{ij}]$. For the hold slack at flip-flop $j$, the range can be calculated similarly as $[\underline{f}+\underline{d}_{ij}, \overline{f}+\underline{d}_{ij}]$. Thereafter, we check all the polygons in the delay model and delete those that are completely outside the bounding box, as illustrated in Figure 7b, where the thick dashed line shows the boundary of the possible region for the flip-flop.

Table 1: Experimental Results of The Piecewise Linear Model and The ILP-based STA

| Circuit | | | Trimming | | Comparison | | | | | | Runtime |
|---|---|---|---|---|---|---|---|---|---|---|---|
| | $n_s$ | $n_g$ | $n_t$ | $g_t$ | $t_s(\%)$ | $t'_s(\%)$ | $v_p^s$ | $v_f^s$ | $v_p^h$ | $v_f^h$ | $T(s)$ |
| systemcdes | 339 | 3617 | 160 | 53 | 1.30 | 1.27 | 875 | 67 | 167 | 10 | 2 |
| wb_dma | 550 | 3780 | 338 | 59 | 2.42 | 2.36 | 3031 | 255 | 177 | 11 | 4 |
| aes_core | 1015 | 26638 | 413 | 54 | 1.04 | 1.02 | 9351 | 246 | 201 | 13 | 11 |
| tv80 | 1044 | 8499 | 433 | 54 | 1.12 | 1.10 | 6199 | 108 | 101 | 11 | 5 |
| mem_ctrl | 2043 | 9833 | 635 | 60 | 0.91 | 0.88 | 1087 | 89 | 152 | 25 | 24 |
| usb_funct | 2262 | 19234 | 763 | 45 | 0.99 | 0.94 | 6123 | 201 | 69 | 11 | 4 |
| ac97_ctrl | 2525 | 11482 | 1328 | 62 | 1.92 | 1.87 | 5580 | 799 | 178 | 12 | 9 |
| pci_bridge32 | 3673 | 16918 | 1620 | 55 | 1.22 | 1.19 | 3918 | 187 | 191 | 10 | 17 |

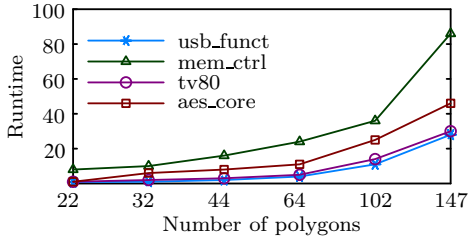

Figure 8: Runtime with different polygon numbers.

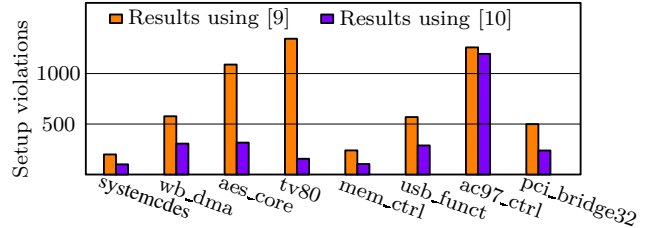

Figure 9: Comparisons of setup violations of flip-flops with [9] and [10]. The proposed method has no violation.

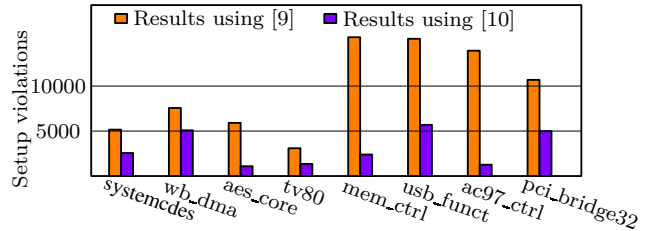

Figure 10: Comparisons of setup time violations of paths with [9] and [10]. The proposed method has no violation.

After trimming polygons, we remove combinational paths and flip-flops that do not need to be included in the ILP formulation. If the source and sink flip-flops of a path always work in the stable region, this path is removed because it does not affect the minimum clock period. If all the paths connected to a flip-flop are removed, the flip-flop is also removed from the ILP formulation to reduce the computational complexity.

## 4 Experimental Results

The proposed framework was implemented in C++ and tested using a 2.67 GHz CPU. We demonstrate the results with circuits from the TAU13 benchmark set. These circuits were synthesized and optimized using a 45nm library and thereafter balanced by clock skews. The number of flip-flops and the number of logic gates in these circuits are shown in the columns $n_s$ and $n_g$ in Table 1, respectively. The interdependency of clock-to-q delay, setup slack and hold slack was characterized with HSPICE. The ILP solver for the optimization problem was Gurobi [13].

In characterizing the three dimensional surface as discussed in Section 3.1, the piecewise model was constructed in 9 minutes and in total 401 points were simulated using HSPICE. The final piecewise model contains 64 polygons. We also simulated the delay surface within the region in which setup and hold slacks are smaller than 100ps with 1ps resolution. The total simulation time of this surface is 2.8 hours, confirming the efficiency of the proposed characterization method.

The results applying the piecewise delay model and the ILP-based timing analysis are shown in Table 1. The number of flip-flops after trimming is shown in the column $n_t$. The number of polygons in the delay model was reduced from 64 to the average number shown in the column $g_t$. These results demonstrate that the trimming technique in our algorithm can reduce the number of flip-flops in the ILP model and the number of polygons effectively, resulting in a much smaller problem space.

To demonstrate the improvement in clock period considering the interdependency of clock-to-q delay, setup slack and hold slack, we compare the clock period calculated by our method with the result from traditional STA. The column $t_s$ in Table 1 shows the relative clock period reduction from STA, in which the setup time is defined as the input slack when the clock-to-q delay is degraded to the 110% of the stable delay. The column $t'_s$ shows the clock period reduction from the result of STA with the setup time defined with respect to the point at which the clock-to-q delay just starts to increase. This setting produces a smaller clock-to-q delay but a larger setup time. In both scenarios, our method achieved up to 2.42% and 2.36% of improvement in the clock period. This reduction of clock period exclusively results from the more accurate modeling and evaluation of the interdependency of clock-to-q delay, setup slack and hold slack, and no additional resource is required. Therefore, the proposed method is very useful in late stages of design flow, where design iteration is normally not preferred.

In traditional STA, the three dimensional clock-to-q delay model is not used. Therefore, STA cannot recover from timing violations. Consider the scenario that the target clock period is the one calculated by the proposed method. With the target clock period given, the column $v_p^s$ in Table 1 shows the number of paths with setup violation in STA, and the column $v_f^s$ shows the number of flip-flops with timing violation. This comparison demonstrates that the proposed method has a significant advantage in removing timing violations. Furthermore, the numbers of paths and flip-flops with hold violation in STA are shown in the columns $v_p^h$ and $v_f^h$, respectively. This comparison confirms again the advantage of the proposed method where there is no timing violation in all these cases.

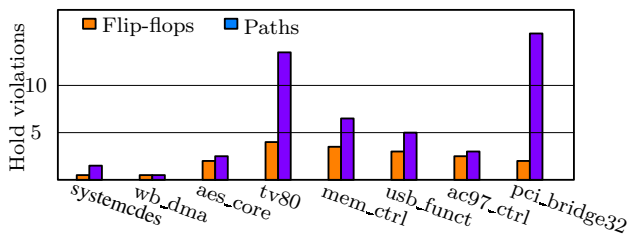

Figure 11: Hold time violations at flip-flops and on paths from [10]. The proposed method has no violation.

The runtime of the proposed method is shown in the last column in Table 1. The largest runtime is 24 seconds for the circuit mem_ctrl with 2043 flip-flops. This runtime is larger than that of a block-based STA algorithm. However, considering that the application scenario of the proposed method is at late stages of design flow, this runtime is acceptable. In the proposed method, the number of polygons in the clock-to-q model affects the runtime of the ILP-based timing analysis. Figure 8 shows the runtime trend of the proposed method with respect to the number of polygons in the delay model. In all these cases, there is no noticeable difference in the calculated clock periods. Generally the runtime increases proportionally to the number of polygons.

To compare the proposed method with previous methods, we first calculate a minimum clock period using our method and use it as the target clock period. Thereafter, we identify the timing violations in the results from the methods in [9] and [10]. Figure 9 and Figure 10 show the comparisons of setup violations of flip-flops and paths, respectively. In [10], only the dependency between setup slack and clock-to-q delay is considered, and it does not exploit the interdependency of clock-to-q delay, setup slack and hold slack together. In the method [9], hold slack is fixed when maximizing the shared setup slack. Consequently, these limitations lead to many timing violations in their results. Besides setup violations, the number of hold violations from the method [10] are shown in Figure 11. This comparison confirms again the effectiveness of the proposed method.

## 5  Conclusion and future work

In this paper, we proposed a holistic method to evaluate the timing performance of a circuit considering the interdependency of clock-to-q delay, setup slack and hold slack. Because this interdependency allows timing compensation across flip-flop stages, the clock period of a circuit can be reduced. This is especially useful in late-stage designs where timing ECO is expensive. The proposed method models the clock-to-q delay surface using a piecewise model, reducing the modeling details by extracting only the necessary delay information useful to timing analysis. Thereafter, the minimum clock period of the circuit is evaluated using an ILP-based formulation, which for the first time provides a holistic solution considering the interdependency to improve circuit performance.

The proposed method uses a piecewise model for the clock-to-q delay and an ILP formulation, so that it is challenging to apply it for yield analysis directly. The future work is to develop a statistical piecewise model for clock-to-q delay under process variations as attempted in [14]. In addition, an analytic statistical timing algorithm should also be considered.